\documentclass[sigconf]{acmart}
\AtBeginDocument{%
  }

\setcopyright{acmlicensed}
\copyrightyear{2018}
\acmYear{2018}
\acmDOI{XXXXXXX.XXXXXXX}
\acmConference[EASE 2025]{The 29th International Conference on Evaluation and Assessment in Software Engineering}{17–20 June, 2025}{Istanbul, Türkiye}
\acmISBN{978-1-4503-XXXX-X/18/06}

\usepackage{package}

\begin{document}


\title{Do Prompt Patterns Affect Code Quality? A First Empirical Assessment of ChatGPT-Generated Code}

\author{Antonio Della Porta}
\email{adellaporta@unisa.it}
\orcid{0000-0003-1860-8404}
\affiliation{%
  \institution{University of Salerno}
  \city{Salerno}
  \country{Italy}
}

\author{Stefano Lambiase}
\email{slambiase@unisa.it}
\orcid{0000-0002-9933-6203}
\affiliation{%
  \institution{University of Salerno}
  \city{Salerno}
  \country{Italy}
}

\author{Fabio Palomba}
\email{fpalomba@unisa.it}
\orcid{0000-0001-9337-5116}
\affiliation{%
  \institution{University of Salerno}
  \city{Salerno}
  \country{Italy}
}

\renewcommand{\shortauthors}{Della Porta et al.}

\begin{abstract}
Large Language Models (LLMs) have rapidly transformed software development, especially in code generation. However, their inconsistent performance, prone to hallucinations and quality issues, complicates program comprehension and hinders maintainability. Research indicates that \emph{prompt engineering}—the practice of designing inputs to direct LLMs toward generating relevant outputs—may help address these challenges. In this regard, researchers have introduced \emph{prompt patterns}, structured templates intended to guide users in formulating their requests. However, the influence of prompt patterns on code quality has yet to be thoroughly investigated. An improved understanding of this relationship would be essential to advancing our collective knowledge on how to effectively use LLMs for code generation, thereby enhancing their understandability in contemporary software development. This paper empirically investigates the impact of prompt patterns on code quality, specifically maintainability, security, and reliability, using the \textsc{Dev-GPT} dataset. 
Results show that Zero-Shot prompting is most common, followed by Zero-Shot with Chain-of-Thought and Few-Shot. Analysis of 7583 code files across quality metrics revealed minimal issues, with Kruskal-Wallis tests indicating no significant differences among patterns, suggesting that \emph{prompt structure may not substantially impact these quality metrics in \textsc{ChatGPT}-assisted code generation}.
\end{abstract}



\maketitle


\section{Introduction}

Generative AI, particularly through Large Language Models (LLMs), is driving significant changes across various domains~\cite{chang2024survey}, including software engineering~\cite{fan2023large}. In this domain, LLMs are transforming workflows by automating code generation, assisting in building software components, aiding in decision-making, and identifying maintenance issues \cite{fan2023large,hou2023large}. As LLM integration expands, researchers are actively studying how these models can support software engineering tasks, documenting their impact on productivity, quality, and problem-solving in development workflows \cite{russo2024_navigating, kumar2024code, liu2024empirical, draxler2023gender, agossah2023llm, khojah2024beyond}.

Although LLMs have demonstrated substantial contributions to software development tasks~\cite{Agarwal2024Copilot, hou2023large}, such as enhancing developers' productivity, they are not without limitations. Their performance often lacks consistency and remains vulnerable to “hallucinations,” referring to the generation of incorrect or irrelevant outputs~\cite{farquhar2024detecting}. Consequently, when employing LLMs for tasks such as code generation, their outputs cannot be assumed accurate or ready for immediate use. Instead, users must rigorously validate the results and frequently engage in prolonged interactions with the tool to achieve the desired outcome. This requirement undermines the initial promise of increased productivity and efficiency, limiting the tool's effective and potentially transformative benefits. Therefore, studying and acquiring foundational knowledge to effectively utilize LLM in the shortest possible time has become essential.


To address these challenges, researchers have increasingly focused on \emph{prompt engineering} defined as \textit{the practice of crafting precise inputs to improve LLM performance by guiding responses without modifying model parameters}~\cite{sasaki2024systematic}. Recent studies have indeed shown that performance may vary by as much as \num{45.48}\% between optimal and suboptimal prompts in some models, underscoring the sensitivity of outputs to prompt design \cite{cao2024worst}. This emphasis on prompt quality has led to the development of \emph{prompt patterns} \cite{white2024chatgpt}, which are structured templates akin to design patterns that offer reproducible frameworks for optimizing LLM output. Examples of these patterns include formats that encourage step-by-step reasoning~\cite{feng2024towards}, define specific personas \cite{krapp2024quasi}, or provide a few illustrative examples \cite{cheng2024novel}, each crafted to enhance different aspects of LLM performance. These patterns have consistently proven effective in producing more reliable, contextually relevant outputs, leading to greater accuracy in code generation \cite{abukhalaf2023codex}.

While current research acknowledges the impact of prompt patterns on code generation, \emph{there is a notable lack of studies investigating how these patterns affect the quality of generated code}. Given the existing body of knowledge, it is reasonable to expect that prompt design could significantly influence aspects such as code maintainability, security, and reliability. However, empirical studies in this area remain limited, partly due to the challenges of constructing datasets that capture real-world usage scenarios. In response to this gap, Xiao et al.~\cite{xiao2024devgpt} recently introduced the \textsc{DevGPT} dataset, a collection of developer interactions with \textsc{ChatGPT}. Later on, Wu et al.~\cite{wu2024chatgpt} conducted a preliminary analysis of prompt patterns with a focus on code quality.

In order to contribute filling this research gap and provide novel knowledge to the field of prompt engineering, we defined the following objective for this investigation.
\steObjectiveBox{\faBullseye\ Paper Objective}{Our objective was to advance understanding of the role of \textbf{prompt patterns} in code generation through an investigation into their relationship with various aspects of \textbf{code quality}, focusing on maintainability, security, and reliability.}

The scientific novelty lies in the analysis of how specific prompt patterns could influence key quality attributes in generated code, with direct implications for software developers.
By examining these dimensions, we provide insights into how prompt engineering can be refined to produce more understandable, secure, and maintainable code, ultimately supporting developers in interpreting and trusting LLM-generated outputs in diverse software development contexts. The prompt patterns analyzed have been informed by the work of Hou et al. \cite{hou2023large} where we selected a set of the prompt patterns that the authors considered the most easy to use.

Moreover, since the dataset used in the work contains only code and conversations that were made using \textsc{ChatGPT}, the scope and implications of this work can't be generalized to all the LLMs. Nonetheless, since \textsc{ChatGPT} is among the most widely used LLM by practitoners in software development \cite{sergeyuk2025using}, our findings impacts a large percentage of practitioners.

Concretely, this study provided three main contributions:
\begin{enumerate}
    \item An \textbf{empirical analysis of prompt patterns and code quality}, revealing no statistically significant relationship between the prompt patterns analyzed and code quality across maintainability, reliability, and security dimensions.

    \smallskip
    \item A \textbf{data contribution} through a refined version of the \textsc{DevGPT} dataset, with duplicates and redundancies removed and enriched with metadata on conversation topics, prompt patterns, and code quality metrics. 

    \smallskip
    \item A \textbf{publicly available online appendix} \cite{appendix}, containing all data and scripts used in this study to support transparency and reproducibility.
\end{enumerate}

\noindent \textbf{Structure of the paper.} Section \ref{sec:related} summarizes the related literature and how our work advances the current state of the art. Section \ref{sec:method} introduces the research questions driving our work and the research methods employed to address them. Section \ref{sec:results}, we present the results of the study, while Section \ref{sec:discussion} provides the actionable implications that our work has for researchers, educators, and practitioners. In Section \ref{sec:ttv}, we discuss the limitations of the work and how we mitigated them while designing the study. Finally, Section \ref{sec:conclusion} concludes the paper and outlines our future research agenda.

\section{Related Work}
\label{sec:related}
Our work builds on the insights from several studies in the realm of Large Language Models for software engineering.

In the context of LLM usage, a prompt is defined as a \textit{``set of instructions or input data provided to a Large Language Model to guide its output''} \cite{ouyang2022training}. Effective prompt design has been shown to directly shape the model's generated responses. For instance, Brown et al.~\cite{brown2020language} demonstrated that \emph{prompt engineering}, that is, the practice of crafting precise inputs to improve the capabilities of LLMs \cite{zhou2022large}, may sometimes yield results comparable to or even better than model fine-tuning, which requires extensive task-specific training data to adjust the model's pre-trained weights. Indeed, prompt engineering leverages the model's existing knowledge, allowing for optimized performance without the need for additional training.

Sasaki et al.~\cite{sasaki2024systematic} further defined \emph{prompt engineering patterns} as \textit{``a systematic approach to structuring interactions, providing a versatile framework applicable across various domains''}. 
The significance of prompt engineering in software engineering has been pointed out by White et al.~\cite{white2024improvingcode}, who found that effective prompt usage can enhance early stages of the software development lifecycle. Similar results were found by Arvidsson and Axell~\cite{arvidsson2023prompt} and Rodriguez et al.~\cite{rodriguez2023prompts}, who assessed the role of prompt engineering on requirements engineering and traceability recovery tasks, respectively. 

Wang et al.~\cite{wang2022no,wang2023prompttuning} examined prompt tuning in various code intelligence tasks, such as defect prediction, code search, code summarization, and code translation, demonstrating that prompt tuning consistently outperforms traditional fine-tuning across these areas. Yu et al.~\cite{yu2024fine} investigated automated code review, showing that refined prompts can significantly improve the accuracy and comprehensibility of code assessments. O'Brien et al.~\cite{obrien2024prompt} studied prompt effectiveness in code generation with \textsc{GitHub Copilot}, analyzing how prompts interact with \textit{TODO} comments to influence code suggestions. Hagar and Masuda~\cite{hagar2024prompt} explored prompt engineering in software testing, finding that customized prompts can assist users of varying expertise, from beginners to experts, in developing effective test architectures. More recent studies continued to expand the applications of prompt engineering in software engineering. Li et al.~\cite{li2024approach} proposed a \textsc{ChatGPT}-based approach for rapid source code development using structured prompts, which improved both the speed and quality of code generation. Fagadau et al.~\cite{fagadau2024analyzing} empirically analyzed the impact of prompt variations on automated method generation with \textsc{GitHub Copilot}, offering insights into how different prompts can affect performance and accuracy in generated methods.

With respect to these previous studies, our work is complementary, as we focus specifically on the impact of prompt patterns on quality attributes. While previous research has demonstrated the effectiveness of prompt engineering in various tasks, including code generation, our work provides a deeper analysis of how prompt patterns may affect maintainability, security, and reliability of the code generated by LLMs. 

A notable contribution to software engineering research was introduced by Xiao et al.~\cite{xiao2024devgpt}, who proposed \textsc{DevGPT}, a dataset specifically designed to investigate how developers interact with \textsc{ChatGPT} in software development contexts. The dataset comprises \num{29778} \textsc{ChatGPT} prompts and responses, including \num{19106} code snippets, and is linked to software development artifacts such as source code, commits, issues, and pull requests. Sourced from shared \textsc{ChatGPT} conversations on \textsc{GitHub} and \textsc{Hacker News}, \textsc{DevGPT} provides a valuable resource for examining developer queries, the effectiveness of \textsc{ChatGPT} for code generation and problem-solving, and its broader impact on software engineering. 


Building upon \textsc{DevGPT}, Wu et al.~\cite{wu2024chatgpt} examined the role of prompt patterns in improving developer-\textsc{ChatGPT} interactions throughout the software development lifecycle. The study focused on (1) analyzing the structure and duration of developer-\textsc{ChatGPT} conversations, (2) identifying prompt patterns that elicit high-quality responses, and (3) optimizing these patterns for software development tasks. As part of this investigation, the authors also evaluated the impact of individual prompt patterns on code quality metrics, assessing responses based on code size, complexity, and nesting levels. Leveraging the prompt patterns proposed by White et al.~\cite{white2023prompt}---which include techniques for enhancing input semantics, customizing outputs, identifying errors, and refining prompts---the researchers assigned quality scores to \textsc{ChatGPT} responses based on an evaluation performed by models such as \textsc{Code-Llama} and \textsc{Mistral}. The study identified multiple patterns, e.g., \emph{Output Customization} and \emph{Error Identification}, as particularly effective, with the former emerging as the most frequently used and impactful pattern, especially in tasks like code generation and software management.

To the best of our knowledge, the work by Wu et al. \cite{wu2024chatgpt} represented the first attempt to investigate the impact of prompt patterns on software quality, making it the closest study to our investigation. When comparing the two studies, multiple aspects should be noted. First, Wu et al. explored a broader range of software engineering tasks, while our study focuses on code generation, enabling a deeper analysis of how prompt patterns may impact multiple quality attributes of the \textsc{ChatGPT}-generated code. Additionally, Wu et al. relied on a catalog of prompt patterns~\cite{white2023prompt} that has yet to be validated by the scientific research community and lacks evidence of adoption by practitioners~\cite{hou2023large}.

In contrast, our study focuses on a set of prompt patterns detailed by Hou et al.  \cite{hou2023large} that are more commonly recognized by practitioners. We specifically selected a set of basic patterns used in a software engineering context, which are also the most easily usable by non-expert practitioners due to their simple structure. 

Finally, since Wu et al.'s initial analysis, the \textsc{DevGPT} dataset has been expanded with new developer-\textsc{ChatGPT} interactions, enabling a more comprehensive exploration of prompt effectiveness.

\steSummaryBox{\faList\ Related Work: Summary and Contribution.}{Existing studies demonstrate that prompt engineering can enhance a range of software engineering tasks, from requirements elicitation to software testing. However, the specific impact of prompt engineering patterns on quality attributes remains underexplored. While recent investigations have offered initial insights, our work delivers a more comprehensive analysis of these quality dimensions.}

\section{Research Design}
\label{sec:method}
The \emph{goal} of this work was to investigate the extent to which specific prompt patterns influence the quality attributes of source code generated by LLMs, with the \emph{purpose} of expanding current knowledge in prompt engineering and potentially uncovering the side effects of code generation on maintainability, reliability, and security. More specifically, our work seeks to verify the following working hypothesis:

\begin{center}
    \textit{The use of different prompt patterns, which are hypothesized to affect the performance of LLMs in generating source code, could lead to variations in the quality of the generated code.}
\end{center}

The hypothesis is supported by recent literature, e.g., \cite{white2024improvingcode,wang2023prompttuning}, which highlights the promising role of prompt patterns in improving model outputs. However, despite their potential, prompt patterns remain a developing area with definitions that lack explicit consideration of quality attributes that are critical to software engineering applications. For instance, current classifications include prompt types like ``Zero-shot'', which provide minimal structural guidance to the model. This limited focus on quality-related aspects differentiates prompt patterns from traditional design patterns, highlighting the need for an empirical investigation that aligns prompt engineering with quality outcomes, thereby directly supporting our hypothesis.


\begin{figure*}
    \centering
    \includegraphics[width=1\linewidth]{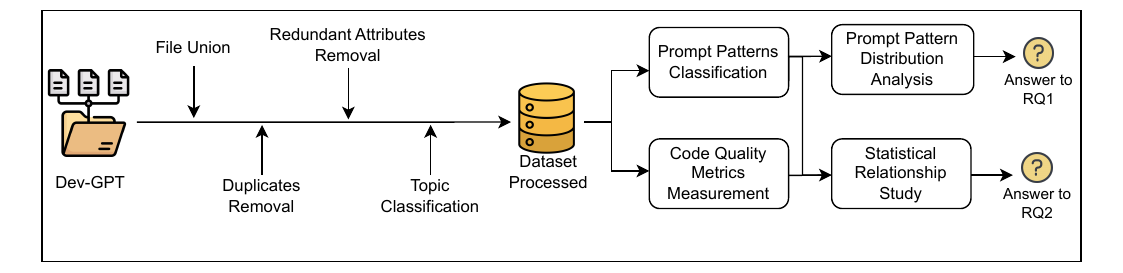}
    \caption{Overview of the Research Method Employed in Our Study.}
    \label{fig:method}
\end{figure*}

\subsection{Research Questions}
To reach the defined goal and test the working hypothesis, we formulated two research questions (\textbf{RQ}s) aiming to shape and guide the research process. In the following, we introduce each research question along with its motivation.

\steResearchQuestionBox{\faQuestionCircle\ \textbf{RQ\textsubscript{1}: Prominence of Prompt Patterns.} \textit{What prompt patterns are most commonly used in conversations with \textsc{ChatGPT}?}}
The first RQ seeks to identify the most commonly used prompt patterns in conversations with \textsc{ChatGPT}. We informed the selection of patterns utilizing the findings of \cite{hou2023large}, specifically selecting a set of basic patterns that practitioners more commonly recognize and use also due to their simple structure. 

The specific choice of \textsc{ChatGPT} is that it is the most widely used LLM by practitioners in software development \cite{sergeyuk2025using}, and so our findings will impact a large percentage of practitioners. This first question must be considered \emph{preliminary}, but \emph{essential} for our research, as understanding the prominence of specific prompt patterns provides a foundational context to interpret the findings of the subsequent analysis. Additionally, this preliminary step may provide insights into the real-world usage of prompt patterns, thereby informing researchers on which patterns are most likely to be applied in practice and guiding future work on optimizing prompt design for practical applications.


\steResearchQuestionBox{\faQuestionCircle\ \textbf{RQ\textsubscript{2}: 
Prompt Patterns and Code Quality.} \textit{Is there a statistical difference in the quality of \textsc{ChatGPT}-generated code when using different prompt patterns?}}
The second research question forms the core of this work, guiding the primary statistical analysis of the potential relationship between different prompt patterns and the quality of generated source code. Addressing this question is essential for achieving the study’s overall objective, as it provides insights into how specific prompt patterns may affect key quality attributes in code. For researchers, these insights expand current knowledge on prompt engineering while offering practitioners practical guidance on selecting prompt patterns to meet quality-related goals in real-world applications. Figure \ref{fig:method} overviews the design of the research procedure employed to address our two research questions. In particular, we adopt multiple steps, as reported in the following:

\begin{enumerate}
    \item First, a dataset of conversations between practitioners and \textsc{ChatGPT} was selected. This dataset, \textsc{DevGPT}, developed during prior research~\cite{xiao2024devgpt}, was deemed relevant for investigating prompt patterns in software development~\cite{wu2024chatgpt}. Various data preparation steps were undertaken, with the most critical being topic classification, which allowed the identification of conversations and prompts relevant to the target users of our investigation, such as developers.

    \item Second, we used LLM-based classification to analyze each conversation. Moreover we also employed a human validation process to ensure the quality of the results. These were applied to classify the prompt pattern used by the developer (addressing \textbf{RQ\textsubscript{1}}).

    \item Third, we used the well-known \textsc{SonarQube} to compute quality metrics (specifically, the number of issues related to maintainability, security, and reliability) of the source code generated by the LLM in response to user requests.

    \item Finally, for each quality metric (our dependent variables), a statistical test was conducted to determine whether there were statistically significant differences in the metrics between the source code generated in response to different prompt patterns (addressing \textbf{RQ\textsubscript{2}}). Since the characteristics of the data, we employed Kruskal-Wallis test as a non-parametric statistical test.
\end{enumerate}

In terms of study design, we followed the guidelines by Wohlin et al. \cite{wohlin2012experimentation}. In terms of reporting, we adhered to the \textit{ACM/SIGSOFT Empirical Standards}\footnote{Available at: \url{https://github.com/acmsigsoft/EmpiricalStandards} };
In particular, we leveraged the \textsl{``General Standard''}, and \textsl{``Repository Mining''} guidelines ensuring to have all the essential attributes and also some desirable attributes. For example, as described in the guidelines we summarized and discussed about the related works and also compared the contributions of the present works in relation to them.

\subsection{Variables of the Study}
The variables of the study were two: the prompt pattern used in the conversation (independent variable) and the quality metrics (dependent variable) of the generated source code.

\begin{description}[leftmargin=0.3cm]
    \item[\underline{Prompt Patterns}.] Regarding the independent variable, it was operationalized using a categorical scale consisting of four prompt pattern categories and their combination, frequently discussed in the literature~\cite{hou2023large}:
\begin{itemize}
    \item \textit{Zero-shot} (ZS) \cite{radford2019language}: It involves providing no examples within the prompt, relying solely on the model's pre-existing knowledge to generate responses.

    \item \textit{Few-shot} (FS) \cite{mann2020language}: It includes a small set of examples within the prompt to help guide the model's understanding and response generation.

    \item \textit{Chain-of-Thought} (CoT) \cite{wei2022chain}: it uses prompts that encourage step-by-step reasoning, aimed at enhancing the model's logical consistency and detail in responses.

    \item \textit{Personas} \cite{kong2024better}: It implements a specific, consistent character or tone in the prompt, fostering contextualized responses aligned with particular roles or perspectives.
\end{itemize}

The selection of these categories was deliberate, as we recognize the existence of additional prompt patterns. First, the decision was made to complement the work of Wu et al.~\cite{wu2024chatgpt}, who relied on a different scale developed during the early phases of LLM development, which has since evolved into the scale used in this study. Second, the chosen patterns are the basic patterns present in literature \cite{hou2023large}, and the most easily usable also by non-expert practitioners due to their simple structure. Given the preliminary nature of this work, we argue that focusing on these prominent patterns is more reasonable than using a broader, yet less representative, set.\footnote{As a methodological note, initially, we attempted to include all identified prompt patterns in our scope. However, due to (1) excessive similarity between some patterns and (2) the limited occurrence of others in the dataset, the results were unsatisfactory. After multiple iterations with different configurations, we decided to focus on the four key prompt patterns reported in the article.}

\item[\underline{Quality Metrics}.] Regarding the dependent variable, we operationalized three proxies of code quality such as (1) \textit{number of maintainability issues}; (2) \emph{number of reliability issues}; and (3) \emph{number of security issues}. The choice of these proxies come from three main considerations. First and foremost, these metrics capture a complementary and broad spectrum of quality attributes, offering insights into the multi-faceted nature of code quality. Indeed, maintenance, reliability, and security provide critical perspectives on how code performs over time, both in terms of robustness and ease of future adaptations \cite{knight2012fundamentals,revilla2007correlations}: as such, their inclusion allowed us to assess software quality in a more comprehensive manner, addressing not only immediate functionality but also the longer-term aspects essential for sustainable and resilient code. In the second place, the combination of these metrics aligns with industry standards, such as \textit{ISO/IEC 25002},\footnote{The ISO/IEC 25002 standard: \url{https://www.iso.org/standard/78175.html}.} and has been widely supported in software quality literature for evaluating factors that affect the sustainability, stability, and security of code over its lifecycle \cite{assal2018security,li2015systematic,tom2013exploration}. Last but not least, these proxies align with practical considerations in the industry, where automated tools like \textsc{SonarQube} are commonly used to assess these attributes~\cite{beller2016analyzing,vassallo2020developers}. 

\end{description}

\subsection{Context of the Study and Data Preparation}

The dataset analyzed in this study, named \textsc{DevGPT}~\cite{xiao2024devgpt}, consists of conversations between developers and \textsc{ChatGPT} focused on various software development tasks. \textsc{DevGPT} was assembled using \textsc{OpenAI}'s conversation-sharing feature. The dataset consists of \num{6} \texttt{json} files representing different data sources used to elicit conversations such as \textsc{GitHub} issues, pull requests, discussions, commits, code files, and threads on \textsc{Hacker News}. To capture the evolving nature of these interactions, data snapshots were taken at multiple intervals, each reflecting the state of the dataset at a particular time. 


The dataset includes separate \texttt{json} files for each source where the conversations had been taken. In total, \textsc{DevGPT} comprises \num{5494} conversation rounds, \num{29788} total prompts containing \num{13988} individual code snippets written across \num{113} different programming languages and frameworks. 

Our initial intention was to use the dataset as-is. However, a preliminary manual review by the first two authors revealed that the dataset contained duplicate conversations and, more critically, included generic conversations that fell outside the scope of software development. For example, one of the user prompts found in the dataset was the following: 

\smallskip
\begin{quote}
    \emph{``Does USB-C without Thunderbolt support two 4k @60Hz monitors?''}
\end{quote}
\smallskip

To ensure that the dataset was reliable, relevant, and representative of software development conversations, we undertook a preliminary data preparation process. We began by merging the various files that make up \textsc{DevGPT}, removing duplicate conversations and unnecessary attributes. Next, we applied a topic classification to filter the conversations, retaining only those that were pertinent to software engineering and development. Given the large volume of prompts, we used a LLM-based classification to expedite this process, relying on \textsc{ChatGPT 4o-mini}.\footnote{For the sake of space limitation, more details on the prompt used in this stage are available in our online appendix~\cite{appendix}.} Since the accuracy of this classification is critical to the quality of our analysis, we also involved \num{10} human experts recruited from our network to validate the LLM-generated classifications. A statistically significant sample of \num{380} prompts was divided among the experts, each of whom reviewed a subset of the prompts using a simple web application we developed (available in the online appendix~\cite{appendix}). This application allowed users to upload a \texttt{json} file containing prompts for sequential review, following the same instructions given to \textsc{ChatGPT}. The experts confirmed all classifications made by the LLM, increasing our confidence in the filtering process used to retain only conversations relevant to software engineering and software development. Consequently, we used this filtered dataset for subsequent analyses.



\subsection{Classification of Prompt Patterns} \label{sec:ppi}

To address the research questions, it is necessary to analyze user interactions with \textsc{ChatGPT}, focusing on identifying and classifying instances of the four prompt patterns selected for the study within these interactions. To classify the prompt patterns in the filtered dataset, we developed an automated LLM-based classification mechanism. As in the data preparation stage, we validated this mechanism through manual analysis to ensure accuracy.


\begin{description}[leftmargin=0.3cm]
    \item[\underline{LLM-Based Prompt Pattern Classification}.] We aimed to use the \textsc{LLAMA} 3.1 70B\footnote{\url{https://www.llama.com/docs/model-cards-and-prompt-formats/llama3\_1}} LLM, but some initial tentative showed that it had difficulties on some very long prompts that the dataset contains. Due to hardware limitations, we were not able to overcome this obstacle with other open-source models, and this led us to opt for \textsc{ChatGPT-4o mini} (snapshot \textit{2024-07-18}). This model was chosen among the others due to a sufficiently large context window to process extensive prompts required for our experimentation. Comparisons were made to \textsc{ChatGPT 4o}, but the results did not diverge in a meaningful way, so we opted for using 4o mini given the lower cost per million tokens of input. 

    To improve the performance of the LLM, we adopted the \textsc{self-refine} approach, as described by Madaan et al. \cite{madaan2024self}, where the model not only produces the initial outcome—in our case, the detection of prompt patterns—but also generates feedback and refines its results based on them.
    To put the \textsc{self-refine} approach in practice, we needed to develop three different prompts for each step of the process\footnote{All the prompts used are available in the online appendix \cite{appendix}.}: the initial classification step, the feedback step, and the refinement step. To increase the results quality and make the process reliable, those prompts were built using \textit{prompt engineering techniques}, like Meta-Prompting \cite{zhang2023meta}, Personas \cite{kong2024better} and Emotion Prompting \cite{li2023large}. After drafting the initial prompt, we conducted an \textit{iterative prompt refinement process}, testing its performance on a small subset of the dataset. This approach ensured that the responses adhered to the specified task while improving the prompt to minimize model hallucinations. A notable adjustment was made to the feedback prompts for classification tasks, which now explicitly constrain the model to reason concisely within a fixed word limit. This change addressed issues in the initial prompts, which were often overly verbose, repetitive, and prone to inferring user intentions that were not explicitly stated. Concretely, consider the following user prompt contained in the dataset (\textit{Conversation\_ID=4, first user prompt}):

\begin{prompt}{User Prompt}
    import cors from 'cors';   
    import axios from 'axios';          
    import fs from 'fs';        
    import express from 'express';   
    import https from 'https';      

    const options = {       
        key: fs.readFileSync('path-to-key'),
        cert: fs.readFileSync('path-to-cert') 
    };         

    const app = express();
    let promptResponse = {};  
\end{prompt}

    In this prompt, the user provides code (we only reported a part of it) and seeks information about its usage (without asking for explicit help). We will take this prompt as an example to comprehend the functioning of the \textsc{self-refine} approach better. For this case, we were performing the task of classifying if a prompt employs a zero-shot or a few-shot prompting pattern. Following the \textsc{self-refine}, the prompt used were:

    \begin{enumerate}
        \item \emph{Initial Classification Prompt}: This generates the initial classification and accompanying reasoning by the LLM. This explicit reasoning is then used in the feedback step to generate constructive feedback. Considering the exemplary prompt, its initial classification was the following:

        \classification{Initial Classification}{\textbf{Classification:} \textit{Few-shot};\\ \textbf{Reasoning:} The prompt consists of multiple code snippets and detailed implementation examples of server-side functionality using Express, Axios, and other libraries. This indicates it is part of a broader instructional context, typical of few-shot prompts that provide examples or detailed instructions to clarify a coding-related query or task.}
        
        \item \emph{Feedback Prompt}: This prompt validates the classification and generates feedback. To reach a convergence to a final classification of all the prompts, this prompt is explicitly written to ask the LLM to write feedback if—when performing the classification task again—it has some suggestions and feedback to provide to the next execution of the task. If the LLM has no feedback or suggestions and so agrees with the classification, it has to return \textit{``None''} as the answer. In the exemplary case, the feedback provided by the LLM was the following:
        
        \classification{Feedback Provided}{\textbf{Feedback:} The classification as “Few-shot” is misleading because the text does not provide indicative examples or different variations of prompts that typically characterize few-shot learning. Instead, it consists of detailed implementation code without direct contextual examples geared for specific tasks.}
        
        \item \emph{Refinement Prompt}: This is used when feedback is generated in the previous step. This prompt is developed to instruct the LLM to perform the classification task again, using the instructions of Step \#1 but with the additional input provided by the feedback. During the refinement process, the LLM produced the following classification for the exemplary case, which is actually the correct one: 
        
        \classification{Refinement Classification}{\textbf{Classification:} \textit{Zero-shot}; \\\textbf{Reasoning:} The prompt consists solely of detailed implementation code without any concrete examples or variations to guide the desired outputs or tasks. There are no explicit instructions or comparative prompting typical of few-shot learning. Thus, it fits the characteristics of a zero-shot prompt, where the task is presented without demonstration.}
    \end{enumerate}
    
    In a preliminary analysis, we observed that the \textsc{self-refine} approach did not consistently converge to a final answer across all elements in the dataset. To address this, we implemented two stopping conditions: (1) when the feedback step produces no new suggestions, and (2) when a fixed number of iterations is reached. During experimentation, we found that by the fifth iteration, the agreement rate between the feedback step and the initial classifications declined significantly, indicating diminishing returns from further iterations. Based on this trend, we set the maximum number of iterations to 5 to optimize the balance between accuracy and efficiency in the classification process.

\smallskip
\item[\underline{Automated Classification Validation}.] 
While the automated classification allowed us to streamline the data collection process, potential misclassifications could impact the reliability of the conclusions of the study. To assess the extent of these misclassifications, we conducted a manual evaluation of the automated mechanism. This evaluation aimed to ensure that the classifications met a high standard of accuracy, addressing the known limitations of LLMs in sometimes producing inaccurate information. For this purpose, two authors acted as \emph{independent inspectors}, each reviewing a representative sample of the classifications to assess their validity. A random sample of \num{10}\% of the total classifications produced by the LLM was selected for evaluation---we deemed this sample sufficiently large to assess the automated classifier. The two inspectors independently analyzed the conversations in the sample, labeling each conversation according to the corresponding prompt pattern. To mitigate potential confirmation bias \cite{oswald2004confirmation}, they conducted their evaluations without prior knowledge of the classifications assigned by the automated mechanism and without any discussion of specific cases.

Following their independent assessments, the two inspectors convened for a focused, two-hour in-person meeting to discuss their findings. During this session, they addressed any disagreements and reached consensus on all classifications. This collaborative step resulted in a manually-curated oracle, which could be compared against the classifications made by the automated classifier. The comparison revealed a \num{97}\% match between the automated and manual classifications, indicating a high level of accuracy in the automated mechanism. This high level of agreement indicated that the automated classification mechanism was largely accurate, providing confidence in the validity of our conclusions and suggesting a limited margin of error.

\end{description}

\subsection{Extraction of Source Code Quality Metrics}
To evaluate the impact of prompt patterns on code quality, we computed software quality metrics for the source code generated by the prompts defined in the dataset using \textsc{SonarQube}. We decided to use \textsc{SonarQube} because (1) the industry uses it since it is a well-known tool used in those contexts to analyze code quality, and (2) it has been used in various software engineering research papers to assess code quality \cite{lenarduzzi2020survey,daragona2023techdebt,lenarduzzi2023critical} as a proof of its reliability. Starting with the source code snippets generated by \textsc{ChatGPT} in the dataset, we created individual code files, assigning file extensions based on the programming language indicated in the dataset. We then executed \textsc{SonarQube} on each file, retrieving results via the \textsc{SonarQube} Web API. The quality metrics obtained for each snippet were recorded in the dataset. For snippets that encountered errors during compilation or analysis, we annotated these cases accordingly to exclude them from the final set of successfully analyzed files, ensuring accurate results.

\subsection{Data Analysis}
To address \textbf{RQ\textsubscript{1}}, we examined the distribution of prompt patterns identified by the automated classifier within the filtered dataset. We measured the frequency of each pattern individually (e.g., only Zero-Shot or only Chain-of-Thought) as well as in combination with other patterns (e.g., Zero-Shot with Chain-of-Thought, Few-Shots with Personas).

To answer \textbf{RQ\textsubscript{2}}, we conducted statistical analyses by formulating the following null and alternative hypotheses:

\begin{itemize}
    \item \textbf{Null Hypothesis:} There are no significant differences in the number of (\textit{H1\textsubscript{0}}) maintainability, (\textit{H2\textsubscript{0}}) reliability, and (\textit{H3\textsubscript{0}}) security issues in the source code generated by \textsc{ChatGPT} based on the prompt pattern used. 
    \item \textbf{Alternative Hypotehsis:} There are significant differences in the number of (\textit{H1\textsubscript{a}}) maintainability, (\textit{H2\textsubscript{a}}) reliability, and (\textit{H3\textsubscript{a}}) security issues in the source code generated by \textsc{ChatGPT} based on the prompt pattern used.
\end{itemize}

To assess differences across multiple groups, we initially selected ANOVA for our analysis. However, as the assumptions for ANOVA were not met, we instead employed the non-parametric Kruskal-Wallis test with Dunn's post hoc test to evaluate the hypotheses, using a significance threshold of $\rho = 0.05$. The statistical analysis was conducted using JASP.\footnote{JASP website: \url{https://jasp-stats.org}}


\section{Analysis of the Results}
\label{sec:results}
Following the experimentation process stated in the previous section, in this section we will delve into the main findings and answer the posed research questions.

\begin{table}
    \centering
    \caption{Ranking of prompt patterns usage.}
    
    \rowcolors{1}{graytable}{white}
    \begin{tabular}{p{0.1\linewidth} p{0.5\linewidth} r}
        \rowcolor{black}
        \textbf{\textcolor{white}{Rank}} & \textbf{\textcolor{white}{Prompt Pattern}} & \textbf{\textcolor{white}{\#Occurrences}}\\
        1 & Zero-shot & \num{10034}\\
        2 & Zero-shot with CoT & \num{713}\\
        3 & Few-shot & \num{576}\\
        4 & Zero-shot with Personas & \num{334}\\
        5 & Few-shot with Personas & \num{134}\\
        6 & Zero-shot with CoT and Personas & \num{107}\\
        7 & Few-shot with CoT & \num{94}\\
        8 & Few-shot with CoT and Personas & \num{49}\\
        \hline
    \end{tabular}
    \label{tab:pp_rank}
\end{table}

\subsection{\textbf{RQ\textsubscript{1}}—Prompt Patterns Prominence}

The complete, filtered dataset included \num{2067} conversations and a total of \num{12045} user prompts directed to \textsc{ChatGPT}. 

Table \ref{tab:pp_rank} presents the final distribution, revealing a strong preference for Zero-Shot prompting, with \num{10034} instances, far surpassing other patterns such as Zero-Shot with Chain-of-Thought (\num{713} instances) and Few-Shot (\num{576} instances). This preference for Zero-Shot prompting likely reflects developers' inclination toward simplicity and efficiency, as it requires minimal setup and leverages the model’s built-in capabilities. 

This finding may be due to various reasons. First, Zero-Shot prompts are highly adaptable across a variety of coding tasks, allowing developers to quickly gauge the model's capabilities without the need for extensive customization, which is particularly beneficial in exploratory phases of development. Second, this pattern has a low barrier to entry, which may be attractive to developers unfamiliar with advanced prompting techniques, as more complex patterns like Chain-of-Thought or Personas require a higher level of understanding and structuring to maximize their effectiveness. Consequently, Zero-Shot prompting may be preferred as it aligns well with the time-efficient, agile workflows often seen in software development, where minimal setup enables rapid iterations and fast prototyping.

Another possible interpretation of this preference is that developers may rely on the model’s pre-trained knowledge and reasoning capabilities to produce quick, actionable insights or initial drafts without extensive guidance, trusting that the model will yield satisfactory results. However, the lower use of advanced patterns, such as Chain-of-Thought or Personas, could suggest an underutilization of techniques that may provide significant advantages in scenarios demanding nuanced reasoning or context-specific responses. Thus, while Zero-Shot prompting remains the default for everyday use, these findings also highlight an opportunity for further training on prompt engineering, which could help developers make better use of advanced patterns to address complex tasks effectively.




\steSummaryBox{\faBarChart\ RQ\textsubscript{1}—Prompt Patterns Presence in Conversation}{The most used prompt pattern in developers conversations is Zero-Shot as shown in Table \ref{tab:pp_rank} with \num{10034} occurrences, Zero-Shot with CoT with \num{713} occurrences and Few-Shot with \num{576} occurrences.}

\subsection{\textbf{RQ\textsubscript{2}}—Prompt Patterns and Code Quality}


\begin{table}
	\centering
	\caption{Descriptive Statistics of the code quality metrics.}
	\label{tab_general_descriptive_statistic}
	{
        \rowcolors{1}{graytable}{white}
		\begin{tabular}{lrrr}
			\hline
            \rowcolor{black}
			\textcolor{white}{\textbf{}} & \textcolor{white}{\textbf{Maintainability}} & \textcolor{white}{\textbf{Reliability}} & \textcolor{white}{\textbf{Security} } \\
			\hline
			\# of Occurrences & $7624$ & $7624$ & $7624$  \\
			Median & $0.000$ & $0.000$ & $0.000$  \\
			Mean & $0.413$ & $0.095$ & $0.002$  \\
			Std. Deviation & $1.855$ & $0.472$ & $0.047$  \\
			Minimum & $0.000$ & $0.000$ & $0.000$  \\
			Maximum & $30.000$ & $12.000$ & $2.000$  \\
			\hline
		\end{tabular}
	}
\end{table}

\begin{table*}
	\centering
	\caption{Descriptive statistics for \textbf{\underline{Maintainability}} issues.}
	\label{tab_mantainability}
	{
        \rowcolors{1}{graytable}{white}
		\begin{tabular}{p{0.15\linewidth}rrrrrrrr}
			\hline
            \rowcolor{black}
			\textcolor{white}{\textbf{MAINTAINABILITY}} & \textcolor{white}{ZS} & \textcolor{white}{ZS-CoT} & \textcolor{white}{FS} & \textcolor{white}{ZS-Personas} & \textcolor{white}{FS-Personas} & \textcolor{white}{ZS-CoT-Personas} & \textcolor{white}{FS-CoT-Personas} & \textcolor{white}{FS-CoT}  \\
			\hline
			\# of Occurrences & $6534$ & $441$ & $390$ & $129$ & $21$ & $32$ & $7$ & $63$  \\
			Median & $0.000$ & $0.000$ & $0.000$ & $0.000$ & $0.000$ & $0.000$ & $1.000$ & $0.000$  \\
			Mean & $0.433$ & $0.263$ & $0.236$ & $0.488$ & $0.190$ & $0.281$ & $1.286$ & $0.365$  \\
			Std. Deviation & $1.953$ & $0.881$ & $0.827$ & $2.020$ & $0.512$ & $0.683$ & $1.604$ & $1.406$  \\
			Minimum & $0.000$ & $0.000$ & $0.000$ & $0.000$ & $0.000$ & $0.000$ & $0.000$ & $0.000$  \\
			Maximum & $30.000$ & $10.000$ & $7.000$ & $14.000$ & $2.000$ & $3.000$ & $4.000$ & $10.000$  \\
			\hline
		\end{tabular}
	}
\end{table*}

\begin{table*}
	\centering
	\caption{Descriptive statistics for \textbf{\underline{Reliability}} issues.}
	\label{tab_reliability}
	{
        \rowcolors{1}{graytable}{white}
		\begin{tabular}{p{0.15\linewidth}rrrrrrrr}
			\hline
            \rowcolor{black}
			\textcolor{white}{\textbf{RELIABILITY}} & \textcolor{white}{ZS} & \textcolor{white}{ZS-CoT} & \textcolor{white}{FS} & \textcolor{white}{ZS-Personas} & \textcolor{white}{FS-Personas} & \textcolor{white}{ZS-CoT-Personas} & \textcolor{white}{FS-CoT-Personas} & \textcolor{white}{FS-CoT}  \\
			\hline
			\# of Occurrences & $6534$ & $441$ & $390$ & $129$ & $21$ & $32$ & $7$ & $63$  \\
			Median & $0.000$ & $0.000$ & $0.000$ & $0.000$ & $0.000$ & $0.000$ & $0.000$ & $0.000$  \\
			Mean & $0.101$ & $0.059$ & $0.051$ & $0.109$ & $0.000$ & $0.125$ & $0.429$ & $0.032$  \\
			Std. Deviation & $0.491$ & $0.318$ & $0.291$ & $0.419$ & $0.000$ & $0.554$ & $1.134$ & $0.177$  \\
			Minimum & $0.000$ & $0.000$ & $0.000$ & $0.000$ & $0.000$ & $0.000$ & $0.000$ & $0.000$  \\
			Maximum & $12.000$ & $3.000$ & $4.000$ & $2.000$ & $0.000$ & $3.000$ & $3.000$ & $1.000$  \\
			\hline
		\end{tabular}
	}
\end{table*}

\begin{table*}
	\centering
	\caption{Descriptive statistics for \textbf{\underline{Security}} issues.}
	\label{tab_security}
	{
        \rowcolors{1}{graytable}{white}
		\begin{tabular}{p{0.15\linewidth}rrrrrrrr}
			\hline
            \rowcolor{black}
			\textcolor{white}{\textbf{SECURITY}} & \textcolor{white}{ZS} & \textcolor{white}{ZS-CoT} & \textcolor{white}{FS} & \textcolor{white}{ZS-Personas} & \textcolor{white}{FS-Personas} & \textcolor{white}{ZS-CoT-Personas} & \textcolor{white}{FS-CoT-Personas} & \textcolor{white}{FS-CoT}  \\
			\hline
			\# of Occurrences & $6534$ & $441$ & $390$ & $129$ & $21$ & $32$ & $7$ & $63$  \\
			Median & $0.000$ & $0.000$ & $0.000$ & $0.000$ & $0.000$ & $0.000$ & $0.000$ & $0.000$  \\
			Mean & $0.002$ & $0.000$ & $0.000$ & $0.000$ & $0.000$ & $0.000$ & $0.286$ & $0.000$  \\
			Std. Deviation & $0.045$ & $0.000$ & $0.000$ & $0.000$ & $0.000$ & $0.000$ & $0.756$ & $0.000$  \\
			Minimum & $0.000$ & $0.000$ & $0.000$ & $0.000$ & $0.000$ & $0.000$ & $0.000$ & $0.000$  \\
			Maximum & $1.000$ & $0.000$ & $0.000$ & $0.000$ & $0.000$ & $0.000$ & $2.000$ & $0.000$  \\
			\hline
		\end{tabular}
	}
\end{table*}

We generated a total of \num{8748} code files from the \textcolor{black}{code snippets contained in the} selected conversations \textcolor{black}{to be evaluated by \textsc{SonarQube}}. Of these, \num{369} Java files were excluded from analysis due to \textsc{SonarQube} license limitations. Out of the remaining \num{8379} files, \num{755} failed to pass the compilation stage and were subsequently excluded from further analysis. To prevent misinterpretation of these uncompiled files as error-free, we annotated their corresponding \texttt{json} entries with a flag, clearly indicating that these files were not analyzed.

The metrics calculated on the final pool of \textcolor{black}{\num{7624}} files are presented in Table \ref{tab_general_descriptive_statistic}. The \textsl{`Security'} dimension was the least frequent in the dataset, with only \num{50} occurrences, indicating few security issues across the analyzed code snippets. In contrast, \textsl{`Maintainability'} had the highest number of issues, totaling \num{3400} across all files, followed by \textsl{`Reliability'}, with \num{750} issues. These overall metrics suggest that maintainability is the most frequently affected quality dimension, which may reflect the general coding practices and limitations of prompt patterns in sustaining code maintainability.

Detailed results by prompt pattern are shown in Tables \ref{tab_mantainability}, \ref{tab_reliability}, and \ref{tab_security}. For maintainability, the ZS configuration shows the highest number of occurrences (\num{6534}), with a mean of \num{0.433} and a substantial standard deviation of \num{1.953}, indicating notable variability in maintainability issues within this pattern. This high variability suggests that while many ZS-generated snippets are maintainable, some exhibit significant maintainability concerns. In contrast, the FS-CoT-Personas configuration, though less frequent (7 occurrences), exhibits the highest mean maintainability issues (\num{1.286}) with a lower standard deviation (\num{1.604}), indicating a smaller but more concentrated set of issues in this pattern.
The median for maintainability remains \num{0} across all configurations, implying that the majority of code snippets are free from maintainability issues, which is consistent across reliability and security metrics. For reliability, mean values are generally low, ranging from \num{0.000} to \num{0.429}, with FS-CoT-Personas showing the highest mean and a slightly elevated standard deviation of \num{1.134}, indicating some inconsistency but generally fewer issues than maintainability. Security issues are particularly sparse, with mean values close to zero for all configurations and a maximum value of \num{2.000} only in FS-CoT-Personas, highlighting the rarity of significant security issues in the considered code snippets.

These distributions indicate that, while the FS-CoT-Personas configuration may correlate with slightly higher incidences of all issue types, most prompt patterns, particularly ZS, result in minimal issues across reliability and security. This analysis suggests that while there are some differences across prompt patterns, the overall impact on code quality metrics is limited, with most configurations yielding generally issue-free code, especially regarding reliability and security.
\begin{table}
	\centering
	\caption{Results of the Kruskal-Wallis statistical test.}
	\label{tab_statistical_test}
	{
        \rowcolors{1}{graytable}{white}
		\begin{tabular}{lrrrl}
			\hline
            \rowcolor{black}
			\textcolor{white}{Fixed Factor} & \textcolor{white}{Statistic} & \textcolor{white}{df} & \textcolor{white}{$\rho$-value} & \textcolor{white}{Rank $\epsilon^2$} \\
			\hline
			Maintainability & $3.801$ & $6$ & $0.704$ & $4.996*10^{-4}$\\
            Reliability & $11.583$ & $6$ & $0.072$ & $0.002$\\
            Security & $2.144$ & $6$ & $0.906$ & $2.818*10^{-4}$\\
			\hline
		\end{tabular}
	}
\end{table}
From a statistical standpoint, two key decisions were made based on the observed distributions: (1) the FS-CoT-Personas configuration was excluded from the analysis due to its low occurrence count, and (2) given the substantial number of outliers, we performed the analysis twice—once including all values and once excluding outliers. This dual approach ensured a thorough and consistent evaluation, as the outliers lacked a discernible pattern or explanation. Notably, the results of the no-outliers analysis mirrored those of the full dataset. The findings of the statistical test, including outliers, are detailed in Table \ref{tab_statistical_test}, with a summary of key results provided below.

\begin{itemize}
    \item \textsl{`Maintainability'}: The test produced a statistic of \num{3.801} with \num{6} degrees of freedom, yielding a $\rho$-value of \num{0.704}. This indicates no statistically significant differences in maintainability across treatments. The effect size (Rank $\epsilon^2$ = $4.996 \times 10^{-4}$) indicates a negligible effect.

    \smallskip
    \item \textsl{`Reliability'}: The test statistic is \num{11.583} with \num{6} degrees of freedom, and a $\rho$-value of \num{0.072}. While this $\rho$-value approaches significance, it is greater than \num{0.05}, indicating no significant difference across treatments. The effect size of \num{0.002} suggests a minimal impact.

    \smallskip
    \item \textsl{`Security'}: The test statistic is \num{2.144} with \num{6} degrees of freedom, with $\rho$-value=\num{0.906}. This indicates a lack of significant differences across treatments. The Rank $\epsilon^2$ of $2.818 \times 10^{-4}$ reflects a negligible effect.
\end{itemize}


\steSummaryBox{\faBarChart\ RQ\textsubscript{2}—Prompt Patterns and Source Code Quality}{The statistical test revealed no statistically significant differences among the prompt patterns for any of the issue variables, resulting in the rejection of all three alternative hypotheses and the confirmation of null hypotheses.}


\section{Discussions and Implications}
\label{sec:discussion}
This section provides a contextual analysis of our findings and highlights potential directions for future research.

\subsection{On the Practitioner's Use of Prompt Patterns}
The results of \textbf{RQ\textsubscript{1}} indicate a clear preference for Zero-Shot prompting among developers, with significantly higher usage than more complex patterns like Few-Shot, Chain-of-Thought, or Personas. As observed in the analysis of the results, this trend may be attributed to the \emph{simplicity, immediacy, and adaptability of Zero-Shot prompting}, which aligns with the efficiency demands of developer workflows, particularly in the exploratory and prototyping phases. By minimizing setup time and requiring no additional context, Zero-Shot prompts allow developers to quickly gauge model capabilities across a wide range of tasks, making it an appealing choice for practitioners focused on speed and agility. Another key factor influencing this preference may be the \emph{current knowledge and familiarity level within the developer community regarding advanced prompt engineering techniques}. As a relatively novel area, prompt engineering is still gaining traction, and many practitioners may be less familiar with patterns that involve higher levels of structuring, such as Chain-of-Thought or Personas. This lack of familiarity, combined with the additional effort required to set up these prompts effectively, may explain why developers gravitate toward Zero-Shot prompts by default, even if other techniques could potentially offer enhanced results for tasks requiring deeper reasoning or contextual awareness. Thus, the findings support previous research indicating that simplicity and usability often drive the adoption of LLM tools in software engineering tasks~\cite{reynolds2021prompt}.

These findings highlight an opportunity for further \emph{dissemination} and \emph{education}. As prompt engineering continues to evolve, increasing awareness of these techniques could enable developers to select patterns more strategically, tailoring them to the complexity and specificity of their tasks. 

\steResearchQuestionBox{\faHandORight\ \textbf{Implication 1}: \textbf{Researchers} should develop a standardized catalog of prompt patterns for code generation tasks, capturing guidelines for their effective use in specific scenarios. \textbf{Educators} can play a key role in disseminating this knowledge by integrating prompt engineering strategies into software engineering curricula.}


\subsection{Prompt Patterns vs Source Code Quality}
In addressing \textbf{RQ\textsubscript{2}}, our analysis reveals that different prompt patterns do not produce statistically significant differences in maintainability, reliability, or security metrics. Although there is some variability in descriptive statistics, particularly within Few-Shot, Chain-of-Thought, and Personas configurations, the Kruskal-Wallis test yielded negligible effect sizes across all metrics, suggesting that prompt patterns alone may not be a decisive factor in influencing code quality.

One explanation for these findings is that \textsc{ChatGPT} are inherently designed to respond effectively to basic prompt structures, making them relatively insensitive to finer adjustments introduced by more structured prompt patterns. Additionally, as prior work highlights~\cite{white2023prompt}, users’ unfamiliarity with advanced prompt engineering techniques may contribute to the limited usage of complex patterns, as observed in \textbf{RQ\textsubscript{1}}. This low adoption of advanced patterns could, in turn, restrict researchers’ ability to fully understand how these patterns might impact code quality, particularly in complex or high-stakes coding tasks. Therefore, further evaluations of complex prompt patterns may be worthwhile, especially as practitioners' familiarity with prompt engineering continues to grow.

\steResearchQuestionBox{\faHandORight\ \textbf{Implication 2}: Findings suggest that \textbf{practitioners} can often achieve satisfactory quality results in code generation tasks using simple prompting techniques, such as Zero-Shot prompting. However, for \textbf{researchers}, this study highlights the need for additional evaluations of complex prompt patterns to better understand their impact on code quality in specialized contexts.}

\subsection{Evaluating Prompt Patterns and Quality Implications in DevGPT Code Generation}
Our findings suggest that most code generated by \textsc{ChatGPT} is relatively free from major quality issues. However, maintainability issues are more prevalent than reliability or security concerns, indicating that \textsc{ChatGPT} may sometimes struggle with structural or stylistic aspects that support maintainable code.

On the one hand, these observations imply that the current set of quality metrics may not fully capture the dimensions of how different prompt patterns impact the quality of generated code. While simple prompts appear to yield satisfactory results for the considered tasks, more complex scenarios may require refined, context-sensitive metrics that better associate prompt patterns with specific quality outcomes. By developing qualitative metrics, researchers could gain a more accurate understanding of how prompt engineering influences code quality, especially in cases where advanced prompt patterns might play a significant role. On the other hand, the dataset itself could benefit from additional diversity in task types and complexity. Expanding \textsc{DevGPT} to include more varied coding scenarios would allow for a deeper evaluation of how advanced prompt patterns function across different contexts. Such diversification would enhance the study of how prompt engineering affect code quality, supporting both routine and complex development needs.

\steResearchQuestionBox{\faHandORight\ \textbf{Implication 3}: The \textbf{research community} should prioritize qualitative analyses and diversified datasets in prompt engineering research. Developing context-sensitive metrics and datasets with varied coding tasks would enable an improved understanding of code quality in LLM outputs, benefiting both standard and specialized software development tasks.}

\section{Threats to Validity}
\label{sec:ttv}
This study presents several threats to validity that we considered as a result of our design~\cite{wohlin2012_experimentation}. 

\smallskip
\textbf{Threats to Internal Validity.} Threats in this category concern the extent to which observed effects can be attributed to the variables studied rather than other factors. In this study, code quality assessments may be affected by the limitations of \textsc{SonarQube}, potentially impacting the consistency of maintainability, reliability, and security metrics. Additionally, biases may be introduced through the iterative self-refinement process with the LLM if feedback loops fail to converge. This risk was mitigated by implementing a maximum iteration limit and conducting manual verification by domain experts.

\smallskip
\textbf{Threats to Construct Validity.} Threats in this category refer to the accuracy with which study measures capture the concepts they intend to represent. In this research, categorizing prompt patterns into specific types—Zero-shot, Few-shot, Chain-of-Thought, and Personas—may oversimplify how prompt patterns affect code generation, as nuances in pattern design and implementation are not fully captured. Patterns were selected based on prevalent literature to address this. Furthermore, employing AI tools to support the classification process may have introduced some inaccuracies, as LLMs are known for occasional hallucinations. However, automatic analysis was supplemented with a manual review conducted by both the authors and external experts, ensuring robust results and mitigating potential risks.

\smallskip
\textbf{Threats to External Validity.} These threats concern the generalizability of findings beyond the study's specific conditions. The use of a single dataset (\textsc{DevGPT}) and a specific LLM model (\textsc{ChatGPT-4o mini}) may limit the applicability of results to other datasets or models. In this respect, future work should consider additional datasets and alternative LLMs to validate the results across different contexts.

\smallskip
\textbf{Threats to Conclusion Validity.} Threats in this category examine whether the statistical analyses accurately reflect relationships between variables. In this study, the Kruskal-Wallis test, a non-parametric alternative to ANOVA, helps to address potential issues related to normality and variance assumptions in the data. However, this test's reliance on rank-based analysis may still introduce limitations, particularly in detecting subtle effects. Care was taken to set significance thresholds to minimize the risk of Type I and II errors, thereby enhancing the robustness of the study's findings.


\section{Conclusions}
\label{sec:conclusion}
This study explored the relationship between prompt patterns and the quality of source code. An automated prompt classification mechanism, powered by \textsc{ChatGPT}, was employed to (1) isolate prompts specifically related to software engineering and (2) identify the presence of four distinct prompt pattern types. The generated code from \textsc{ChatGPT}, as documented in the \textsc{DevGPT} dataset, was assessed for maintainability, reliability, and security using the widely recognized tool \textsc{SonarQube}. The collected data supported an empirical analysis to investigate the correlation between prompt patterns and the aforementioned code quality metrics. The results indicated that there is no statistically significant relationship between the analyzed prompt patterns and code quality across the evaluated dimensions.
As part of our future research we aim to deepen the analysis by considering more dimensions of code quality such as functional correctness or code smells.

\section*{Acknowledgment}
This work has been partially supported by the \textsl{Qual-AI} national research project, which has been funded by the MUR under the PRIN 2022 program (Code: D53D23008570006). In addition, the work was partially supported by the \textsl{AIMS} project, which is funded by the Fondazione FAIR – Future Artificial Intelligence Research, a non-profit organization established to implement interventions financed under the Italian National Recovery and Resilience Plan. 

\balance
\bibliographystyle{ACM-Reference-Format}
\bibliography{references}

\end{document}